\documentclass[12pt]{article}

\usepackage{amsmath,amsfonts,amssymb,amsthm,amstext,amscd,eucal}
\usepackage{bbold}
\usepackage{array}
\usepackage[latin1]{inputenc}
\usepackage[all]{xy}
\usepackage{hyperref}

\usepackage{latexsym}
\usepackage{graphicx}
\input epsf.sty

\newcommand{\beq}{\begin{equation}}
\newcommand{\eeq}{\end{equation}}
\newcommand{\eqn}{\begin{eqnarray}}
\newcommand{\enn}{\end{eqnarray}}
\DeclareMathOperator{\AdS}{AdS}
\DeclareMathOperator{\dvol}{dvol}
\newcommand{\RR}{\mathbb{R}}
\newcommand{\ZZ}{\mathbb{Z}}
\newcommand{\CC}{\mathbb{C}}
\DeclareMathOperator{\Sph}{S}
\newcommand{\SO}{\mathrm{SO}}

\newcommand{\fso}{\mathfrak{so}}
\newcommand{\be}{\boldsymbol{e}}

\def\Label#1{\label{#1}%
  \smash{\hbox to0pt{\raise1ex\hbox{\tiny[#1]}\hss}}}
\def\noLabels{\let\Label=\label}
\def\nobbibitem{\let\bbibitem=\bibitem}

\def\CC{{\cal C}}

\def\CM{{\cal M}}
\def\CN{{\cal N}}

\newcommand{\bbibitem}[1]{\bibitem{#1}\marginpar{#1}}

\newcommand{\ads}[1]{{\rm AdS}_{#1}}

\newcommand{\bea}{\begin{eqnarray}}
\newcommand{\eea}{\end{eqnarray}}

\begin{document}

\renewcommand{\thepage}{\arabic{page}}
\setcounter{page}{1}
\noLabels 
\nobbibitem 

\rightline{UPR-1110-T, hep-th/0502111}

\vskip 1cm \centerline{\large \bf Much Ado About Nothing} \vskip
1cm

\renewcommand{\thefootnote}{\fnsymbol{footnote}}
\centerline{{\bf Vijay
Balasubramanian\footnote{vijay@physics.upenn.edu}, 
Klaus Larjo\footnote{klarjo@physics.upenn.edu} and
Joan Sim\'on\footnote{jsimon@bokchoy.hep.upenn.edu} }} 
\vskip .5cm
\centerline{\it David Rittenhouse Laboratories, University of
Pennsylvania} \centerline{\it Philadelphia, PA 19104, U.S.A.}

\setcounter{footnote}{0}
\renewcommand{\thefootnote}{\arabic{footnote}}

\begin{abstract}
We describe the semiclassical decay of a class of orbifolds of AdS space via a bubble of nothing.   The  bounce  is the small Euclidean AdS-Schwarzschild solution.  The negative cosmological constant introduces subtle features in the conservation of energy during the decay.   A near-horizon limit of D3-branes in the Milne orbifold spacetime gives rise to our false vacuum.  Conversely, a focusing limit in the latter produces flat space compactified on a circle.  The dual field theory description involves a novel analytic continuation of the thermal partition function of Yang-Mills theory on a three-sphere times a circle.
\end{abstract}


\section{Introduction}

Flat space with one circular direction around which fermions are anti-periodic (thus breaking supersymmetry) is unstable to decay to {\it nothing} \cite{perry, wittenbubble}.   The space is literally eaten up by an expanding bubble.   In this paper we show how such catastrophes can occur in universes with  a negative cosmological constant ($\Lambda$).   Specifically, we show that a certain analogue of the flat space Kaluza-Klein vacuum, namely AdS space with a circle in it, decays via a bubble of nothing.\footnote{The relevant bubble of nothing solutions have been described in \cite{birm,vijaysimon}.  Charged bubbles have appeared in~\cite{biswas}.   The role of such instabilities in string theory, particularly in the context of a landscape of vacua has been explored in~\cite{dine}.   Properties of  bubbles of nothing in diverse situations are studied in \cite{clean,simongeorgina}.}  This raises the possibility that a highly non-perturbative and ill-understood process in gravity might have a simple explanation as barrier penetration in a dual field theory effective potential.

In Sec.~2 we describe our false vacuum from three different perspectives.   First, the space is a so-called ``topological black hole'' constructed by identifying global $\ads{5}$ space along a boost \cite{banados}.    This non-supersymmetric orbifold is the analogue in five dimensions of a BTZ black hole \cite{BTZ} in three dimensions.  We suggest an interpretation of the entropy of the black hole in terms of the dual boundary field theory and also compute the mass of the spacetime.  We show that it can also be understood as a near-horizon limit of flat D3-branes filling the Milne spacetime, and as an alternative analytic continuation of global AdS space.   We also demonstrate that focusing on the geometry near the horizon of this geometry leads to the flat space Kaluza-Klein vacuum, while focusing on the region between the singularity and the horizon produces the Milne spacetime again.

In Sec.~3 we demonstrate that the AdS bubble of nothing described in \cite{birm,vijaysimon} mediates the decay of the topological black hole.    The bounce solution is simply the Euclidean Schwarzschild black hole.   Famously, there are two such black holes at any temperature exceeding a certain bound,  roughly speaking one larger and one small than the AdS scale.   The minimum temperature translates in our setting into absolute stability of topological black holes with a circle that exceeds a certain size.   Further, only the smaller Euclidean black hole has a non-conformal negative mode and hence this provides the required bounce.   A semiclassical decay of this kind requires conservation of energy.  Because the Lorentzian spacetimes in our setting are time-dependent, total energy is not conserved, but we are nevertheless able to show that in order for the decay to happen the nucleated bubble will have to be accompanied by a bath of energy making up the instantaneous mass difference between the topological black hole and its decay product.  This is an important difference compared to the decay of the flat space Kaluza-Klein vacuum and arises because of the peculiar properties of space with a negative cosmological constant.

 While our discussion is complete as a  description of the decay of five dimensional spaces with $\Lambda < 0$, in string theory the presence of an additional $S^5$ compactification manifold introduces a subtlety.  In this context, we explain that it is expected on entropic grounds that the small Schwarzschild black hole localizes in the $S^5$ via a perturbative instability.  In the flat space focusing limit described in Sec.~2, this perturbative instability is essential.  In this limit, the complete spacetime becomes $R^{1,8} \times S^1$, which decays \cite{wittenbubble} via an eight dimensional bubble of nothing.   Our bounce would lead to a four dimensional hole in spacetime which is spread out evenly over the remaining four non-compact dimensions.  The localization instability is precisely what is required to produce the correct minimum action instanton.

The decay that we describe should manifest itself as a tunneling process in a suitable effective potential in the dual field theory living on the boundary which is here three-dimensional de Sitter space times a circle (with anti-periodic boundary conditions on the fermions).   The flat space limit that we describe should give a field theory realization of the classic instability of the Kaluza-Klein spacetime.     The basic challenge, which we have not solved, is to compute the appropriate effective potential, particularly given the large 't Hooft coupling and the lack of supersymmetry.   Because the Euclidean spacetimes in question are simply thermal AdS and the Schwarzschild-AdS black holes, what we seek is simply a novel analytic continuation of the effective potential of thermal $\CN =4$ Yang-Mills on a sphere which also describes the famous Hawking-Page transition.  Here we restrict ourselves to comments on how some known results in the literature may apply.

We also describe a much more general class of AdS orbifolds which are related to flat space fluxbranes, and describe how these should also decay to nothing.   Brief appendices describe how flat limits of AdS quotients are obtained,  and brane probes of our spacetimes.

\section{The false vacuum}
 \label{sec:review}

The flat space Kaluza-Klein vacuum ($R^{3,1} \times S^1$) that decays to nothing in \cite{wittenbubble}  can be thought of as an orbifold of $R^{4,1}$.   Thus a natural candidate for a false vacuum undergoing such a process with $\Lambda < 0$ would be a non-supersymmetric orbifold of $\ads{5}$ that creates a non-contractible circle. We study one such orbifold, the topological black hole of \cite{banados}.   As we will see, there is a flat space limit of this space that is precisely  $R^{3,1} \times S^1$, making it a natural candidate for a decay to nothing.

\subsection{The topological black hole as an orbifold}
The five dimensional topological black hole studied in~\cite{banados} is an orbifold of AdS space obtained by identification of points  by the action of the Killing vector
\begin{equation}
  \xi = \frac{r_+}{R_{\AdS}}(x_4 \partial_5 + x_5\partial_4)~, 
 \Label{eq:boost}
\end{equation}
where one describes AdS as the universal covering of the surface
\begin{equation}
  -x_0^2 + x_1^2 + x_2^2 + x_3^2 + x_4^2 - x_5^2 = -R_{\AdS}^2
  \Label{AdShyper}
\end{equation}
in $\mathbb{R}^{2,4}$.   Since the norm of the Killing vector can be negative, the resulting spacetime has closed timelike curves \cite{banados,paper5.2}.   These regions are excised, leading to a spacetime that is interpreted as a black hole in precise analogy to the BTZ black hole in three dimensions \cite{BTZ}.      The null hypersurfaces separating the regions of positive and negative norm of $\xi$ become singularities as timelike geodesics end there in finite proper time.   Thus the curvature is locally trivial, but the global causal structure is that of a black hole (see Fig.~1 and~\cite{banados}).

\begin{figure}
\begin{center} 
\epsfysize=8cm
\epsfbox{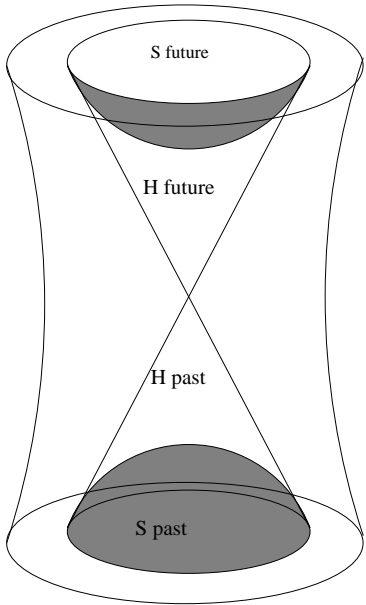}\label{f1}
\caption{The topological black hole.   The surfaces $S_{f,p}$ are future and past singularities where timelike geodesics end.   Infinity is connected and is topologically $S^3 \times S^1 \times R$ where $R$ represents time.   $H_f$ is a future horizon -- light rays from the region interior to $H_f$ cannot reach infinity. (For details on the causal structure, see~\cite{banados,simongeorgina}).  }
\end{center}
\end{figure}

Specifically, the singularities occur at the surfaces $\xi^2 = 0$, which are represented by the two disconnected branches of a hyperboloid:
\begin{eqnarray}
S_f & : & x_0 = \sqrt{R_{\AdS}^2 + x_1^2 + x_2^2 + x_3^2}, \\
S_p & : & x_0 = -\sqrt{R_{\AdS}^2 + x_1^2 + x_2^2 + x_3^2}.
\end{eqnarray}
The Killing vector $\xi$ is timelike in the causal future of $S_f$
and in the causal past of $S_p$, and it is these regions which will contain closed timelike curves and are removed.  The horizons occur on the surface $\xi^2 = r_+^2$, which gives rise to two null cones,
connected at the apex:
\begin{eqnarray}
\label{hor}
H_f & : & x_0 = \sqrt{x_1^2 + x_2^2 + x_3^2}, \\
H_p & : & x_0 = -\sqrt{x_1^2 + x_2^2 + x_3^2}.
\end{eqnarray}
$H_f$ is a future horizon, but because infinity is connected the lower
cone does not represent a true horizon (see the Penrose diagrams in \cite{banados} and discussions in \cite{simongeorgina}).

A  global description of the topological black hole is achieved by parametrizing global AdS space (\ref{AdShyper}) as \cite{banados}:
\begin{equation}
  \begin{aligned}[m]
    x^\mu & =  \frac{2R_{\AdS}}{1 - y^2}\, y^\mu~, \quad
    \mu = 0,1,2,3 \\
    x_4 & = r\, \frac{R_{\AdS}}{r_+} \sinh \frac{r_+}{R_{\AdS}}\chi~, \\
    x_5 & =  r\,\frac{R_{\AdS}}{r_+} \cosh \frac{r_+}{R_{\AdS}}\chi~, \\
     r & = r_+ \frac{1+y^2}{1-y^2}~,
  \end{aligned}
 \Label{eq:kruskal}
\end{equation}
where $y^2 = y^\mu\,y^\nu\,\eta_{\mu\nu}$ is the lorentzian norm
and all $\{y^\mu\}$ are non-compact coordinates subject to
the constraint $-1 < y^2 < 1$.   In these coordinates the AdS metric is
\begin{equation}
  ds^2 = \frac{R_{\AdS}^2}{r_+^2}\,(r+r_+)^2\,dy^\mu\,dy^\nu\,\eta_{\mu\nu}
  + r^2 d\chi^2~,
 \Label{eq:topbh}
\end{equation}
while the Killing vector of the identification is $\xi = \partial_{\chi}$  Thus, the topological black
hole arises by identifying $\chi \sim \chi + 2\pi$ in (\ref{eq:topbh}).   This Kruskal-like coordinate system covers the entire spacetime between the connected boundary at infinity and the singularities.

In this coordinate system $r \to r_+$ at the horizon, where the proper size of the $\chi$ circle becomes $2\pi r_+$, giving a physical meaning to the orbifold identification parameter in (\ref{eq:boost}).  The conformal boundary of the spacetime is reached as $y^2 \to 1$.  The corresponding equation, $1=y^2 = y^\mu\,y^\nu\,\eta_{\mu\nu}$, defines a unit curvature three dimensional de Sitter spacetime.    Thus (\ref{eq:topbh}) shows that the conformal boundary of the topological black hole is three dimensional de Sitter
space times a circle, $dS^3\times S^1$:
\begin{equation}
  g_\partial = \frac{R_{\AdS}^2}{r_+^2}\,g_{dS^3} + d\chi^2~.
 \label{eq:bmetric}
\end{equation}
We have conformally rescaled the boundary metric by $r^2$.  The ratio of the AdS scale to $r_+$ will be  a physical scale parameter in the dual field theory.  The area of the horizon at $y^2 = 0$ is
\begin{equation}
A_{\textrm{Horizon}} = 
32 \pi^2 R_{\AdS}^2 r_+ y_0^2
\Label{growhorizon}
\end{equation}
Unusually, this grows with time $y_0>0$.   A possible interpretation in the dual field theory is as follows.  Since the dual is defined on de Sitter space, the expansion of the geometry will constantly stretch modes from the UV into the deep infrared.  Hence, local CFT experiments done within a finite sized box will have access to fewer and fewer modes as time passes.    Given the usual AdS/CFT relation between the deep interior of spacetime and the deep infrared of the dual theory, the stretching of CFT modes by the geometry should correspond to the increasing horizon area.   An alternative interpretation is given by Ross and Titchener who argue that the meaningful entropy relation here is actually between the cosmological horizon seen by the inertial boundary observer and an analogous construct in the bulk \cite{simongeorgina}.

Another convenient coordinate system that covers only the exterior region of the topological black hole ($0\leq y^2\leq 1$) is \cite{banados}:
\begin{equation}
ds^2 = \frac{R_{{\rm Ads}}^2}{r^2-r_+^2} dr^2 + (r^2-r_+^2)\left[
  -dt^2 + \frac{R_{{\rm Ads}}^2}{r_+^2}
  \cosh^2 \frac{r_+t}{R_{{\rm Ads}}} d\Omega_2 \right] + r^2 d\chi^2~.
\Label{schwarzschild}
\end{equation}
This ``Schwarzschild'' coordinate system has the attractive feature that it foliates the spacetime with equal radius slices that are conformal to de Sitter space times a circle.   Hence the $dS_3 \times S^1$ structure of the conformal boundary at $r \to \infty$ is evident.

Finally, de Sitter space can be written in coordinates appropriate to a static observer in this accelerating geometry (see \cite{accreview} for a review).    After rescaling time by $r_+/R$ and transforming to such static coordinates the topological black hole metric becomes
\begin{equation}
ds^2 = \frac{R_{{\rm Ads}}^2}{r^2-r_+^2} dr^2 + (r^2-r_+^2) {R_{{\rm Ads}}^2 \over r_+^2} \left(
-(1-\rho^2) dt^2 + \frac{1}{1-\rho^2} d\rho^2 + \rho^2
d\theta^2\right) +   r^2 d\chi^2
\Label{staticmet}
\end{equation}
where $0\le\rho<1$ and $0\le\theta\le 2\pi$.   This static metric does not cover the full exterior of the topological black hole.   On the boundary it is the metric appropriate to an inertial boundary observer.  Note the cosmological horizon seen by such an observer is at $\rho = 1$.    The corresponding entropy is related to a bulk horizon area in \cite{simongeorgina}.

It is straightforward to embed the above discussion in Type IIB string theory. Since the topological black hole is locally $AdS_5$, its direct product with a five sphere, along with a constant
dilaton and self-dual five form Ramond-Ramond flux solves the type IIB
equations of motion.  (Appendix B constructs the relevant flux while exploring brane probes of this spacetime.)   Following the analysis in \cite{paper1,paper5.1}, the boost orbifold (\ref{eq:boost}) breaks all supersymmetry.     In addition, the $S^1$ admits both periodic and antiperiodic boundary conditions for fermions.  We will be interested in the anti-periodic case, because we will only find an instanton for tunneling to nothing in this situation.

\subsection{Branes on the Milne spacetime}

The topological black hole can also be obtained as a near horizon limit of N D3-branes filling the boost orbifold $\RR^{1,1}/\ZZ$.   The metric describing N coincident D3-branes is
\begin{equation*}
  g = H^{-1/2}\,ds^2(\RR^{1,3}) + H^{1/2}\,ds^2(\RR^6)~,
\end{equation*}
where $H$ is a harmonic function in $\RR^6$.    Orbifolding by a boost on $\RR^{1,3}$ does not change this metric locally.  Thus   N D3-branes filling $\RR^{1,1}/\ZZ$ are described by
\begin{equation*}
  g = H^{-1/2}\,\left[ds^2(\RR^{1,1}/\ZZ) + ds^2(\RR^2)\right] +
  H^{1/2}\,ds^2(\RR^6)~.
\end{equation*}
By construction, the near horizon geometry of this space will be a boost 
orbifold of the Poincar\'{e} patch of $\ads{5} \times S^5$:
\begin{equation}
  g = \frac{r^2}{R_{{\rm AdS}}^2}\,\left[ds^2(\RR^{1,1}/\ZZ) + ds^2(\RR^2)\right] +
  \frac{R_{{\rm AdS}}^2}{r^2}\,dr^2 + R_{{\rm AdS}}^2\,g_{\Sph^{5}}~.
  \Label{poincareorb}
\end{equation}
Following the discussion in the appendix of \cite{paper5.2} based on
\cite{joanAdS}, this can be extended to an orbifold of global AdS space.\footnote{In flat space all boosts are equivalent, but in AdS space this is not precisely true.  This is why the results of \cite{joanAdS,paper5.2,simonowen} are needed here.}  By comparing the generator of the orbifold identifications, it is easy to show that the global spacetime is precisely the topological black hole.

The patch (\ref{poincareorb}) covers an  unusual region of the topological black hole that coincides neither with the Kruskal nor  the Schwarzschild regions described above.   Specifically, 
the quotient $\RR^{1,1}/\ZZ$ is described by two charts
\begin{equation}
  ds^2(\RR^{1,1}/\ZZ) = -d\chi^2 + \chi^2\,d\psi^2 \quad \quad t^2\,>\,x^2~,
\end{equation}
where $\{t,\,x\}$ stand for the cartesian coordinates in $\RR^{1,1}$,
and by
\begin{equation}
  ds^2(\RR^{1,1}/\ZZ)= d\chi^2 - \chi^2\,d\psi^2 \quad \quad x^2\,>\, t^2~.
\end{equation}
This second chart covers a region with closed timelike curves  since
$\psi\sim\psi + 2\pi$ after the orbifold identification takes place. Therefore it is excised from the spacetime to give rise to the topological black hole.

Since our spacetime arises as the near-horizon limit of a stack of D-branes whose worldvolume is identified, the dual is expected to be precisely an $\CN = 4$ SU(N) Yang-Mills theory on the boundary.\footnote{Since open strings do not have twisted sectors we do not expect additional degrees of freedom in the theory and since the transverse space is left alone, there will not be any image D-branes to deal with.  Supersymmetry will be broken  by the boundary geometry, curvature couplings and  fermion periodicity.}  The Poincar\'e patch region that arises naturally from the near horizon construction has a boundary which is Milne space times a circle.  This will be conformal to a patch of the ${\rm dS}_3 \times S^1$ global boundary.

\subsection{Euclidean continuation}

Another interesting feature of the topological black hole is that its  euclidean continuation is simply thermal AdS (this was also observed previously in \cite{cai}).  This will imply a connection between thermal $\CN=4$ Yang-Mills and the physics of the decay of the topological black hole.

We can continue the Kruskal metric (\ref{eq:topbh}) to Euclidean signature as $y^0\,\to\,ix$.   This automatically maps the ${\rm dS}_3 \times S^1$ Lorentzian boundary to  $S^1\times S^3$ since the Euclidean section of de Sitter space is simply a round sphere.  The metric for the euclidean topological black hole becomes
\begin{equation}
  g_{\text{Etop}} = (2R_{\text{AdS}})^2\,
  \frac{d\vec{x}^2}{(1-\vec{x}^2)^2} + r_+^2\,
  \frac{(1+\vec{x}^2)^2}{(1-\vec{x}^2)^2}\,d\chi^2~.
 \label{eq:etop}
\end{equation}
where we replaced the $y^{1,2,3}$ in (\ref{eq:topbh}) by $x^{1,2,3}$.    Now change coordinates as
\begin{equation*}
  \cosh\rho = \frac{1+\vec{x}^2}{1-\vec{x}^2} \quad ,\quad
  x^i = (\cosh\rho)^{1/2}\,\hat{x}^i\,,
\end{equation*}
where $\{\hat{x}^i\}$ parameterise a 3-sphere.  In these coordinates \eqref{eq:etop} becomes 
\begin{equation}
  g_{\text{Etop}} = R_{\text{AdS}}^2\,\left(d\rho^2 +
  \sinh^2\rho\,g_{S^3}\right) + r_+^2\,\cosh^2\rho\,d\chi^2~.
 \Label{eq:etopbis}
\end{equation}
This is precisely thermal euclidean AdS with
$r_+ \equiv R_{\text{AdS}}\,\beta$, where $\beta$ is the
standard $1/T$ period in thermal physics. Thus, temperature in the thermal AdS interpretation
is identified with the inverse of the radius of the circle in the topological black hole evaluated at the horizon.

Going the other way, we can recover the Schwarzschild coordinates for the topological black hole by a novel analytic continuation of thermal AdS.   Analytically continue an azimuthal angle ($\theta \to i \tau$) in (\ref{eq:etopbis}).   The 3-sphere in thermal AdS then becomes 3d de Sitter space and the overall metric is
\begin{equation}
  g = r_+^2\,\cosh^2\rho\,d\chi^2 + R_{\text{AdS}}^2\,
  \left(d\rho^2 + \sinh^2\rho\,g_{\text{dS}^3}\right)~.
  \Label{scont}
\end{equation}
The further transformation $r = r_+ \cosh\rho$ combined with the rescaling $t = \tau R/r_+$ yields the Schwarzschild coordinates (\ref{schwarzschild}) for the topological black hole.

\subsection{Focusing limits}

Given a locally AdS space one can construct limits that focus onto small regions of the geometry to arrive at locally flat spacetimes.   Such flat limits of global AdS space always lead to global Minkowski space.    Flat limits of AdS quotients can, however, lead to different results.  Mathematically, a focusing limit results in a contraction of the symmetry algebra and the different contractions lead to different locally flat spacetimes.  This is discussed in greater detail in Appendix A.   Using these results it can be shown that the topological black hole \eqref{eq:topbh} must have two inequivalent flat limits.   This is because, using Appendix A,  the boost generator \eqref{eq:boost} belonging to $\fso(2,4)$ can give rise to
either a boost in $\fso(1,3) \ltimes \RR^{1,3}$ or to a
translation belonging to the same algebra, in the flat limit 
$R_{\AdS}\to\infty$.  Below we will construct these focusing limits explicitly in terms of the spacetime metric and show that one leads to the Milne orbifold, and the other to the 5d Kaluza-Klein vacuum.  The latter limit indicates that at least locally the spacetime admits an instability to nucleate a bubble of nothing.

\subsubsection{Kruskal coordinates}
In Kruskal coordinates  \eqref{eq:topbh} consider the scaling limit
\begin{equation}
  R_{\AdS} \to \infty~~~;~~~ \ \ y^\mu \to \frac{\tilde{y}^\mu}{2R_{\AdS}}~~~;~~~
  \ \ r_+~, \tilde{y}^\mu~,\chi\,\, \textrm{fixed}
 \label{eq:flatkk}
\end{equation}
(The second of these equations should be understood as a coordinate transformation followed by a scaling of $R_{\AdS}$.)
The metric becomes that of the Kaluza-Klein vacuum\footnote{The
scaling limit on the $S^5$ in the Type IIB geometry leads to $R^5$ in the standard way.}
\begin{equation*}
  g \to g(\RR^{1,3}) + r_+^2\,d\chi^2~.
\end{equation*}
In this scaling limit $r \to r_+$; so it can interpreted as focusing onto the horizon of the topological black hole.  Thus, the geometry close to the horizon is the Kaluza-Klein vacuum.  If we choose anti-periodic boundary conditions for fermions around the circle in the geometry, we should (at least locally) expect a semiclassical instability \cite{wittenbubble}.

An alternative flat limit breaks the manifest Lorentz covariance in Kruskal coordinates
\eqref{eq:kruskal}.  That is, $\{y^\mu\}=\{y^0,\,\vec{y}\}$ scale
differently in the limit $R_{\AdS}\to\infty$:
\begin{equation}
  \begin{aligned}[m]
    R_{\AdS}\to\infty~, \quad & r_+\to\infty~, \quad \frac{r_+}{R_{\AdS}}
    \quad \text{finite} \\
    y^0 \to 1 - \frac{t}{R_{\AdS}}~, & \quad 
    \vec{y} \to \frac{\vec{z}}{R_{\AdS}}~~~;~~~ t , \vec{z} \ \ {\rm fixed}
  \end{aligned}
 \label{eq:flatmilne}
\end{equation}
Since $y^0 \to 1$ and $\vec{y} \to 0$ in this limit, we can interpret this as focusing on the vicinity of the singularity.  
The scaled metric is
\begin{equation}
  g \to -dt^2 + d\vec{y}^2 +
  \left(\frac{r_+}{R_{\AdS}}\,t\right)^2\,d\phi^2~,
 \label{eq:milne}
\end{equation}
which describes the Milne  boost orbifold $\RR^{1,1}/\ZZ$, in the region
where there are no closed timelike curves.

\subsubsection{Schwarzschild coordinates}

It is useful to present the focusing limit that yields the Kaluza-Klein vacuum in Schwarzschild coordinates (\ref{schwarzschild}).    Make the coordinate transformations
\begin{equation}
 r = r_+ \cosh\frac{\rho}{R_{{\rm AdS}}} ~~~~;~~~~
 \tau = \frac{r_+t}{R_{{\rm AdS}}}
 \end{equation}
in (\ref{schwarzschild}) and take the limit $R_{{\rm AdS}} \to \infty$ while holding $r_+, t$ and $\chi$ fixed.    The metric becomes
\begin{equation}
ds^2 = d\rho^2 - \rho^2 d\tau^2 + \rho^2 \cosh^2 \tau d\Omega_2 + r_+^2 d\chi^2.
\end{equation}
The further change of coordinates $\tilde{r} = \rho \cosh \tau, \quad \tilde{t} = \rho \sinh \tau$
gives the metric $ds^2 = -d\tilde{t}^2 + d\tilde{r}^2 + \tilde{r}^2 d\Omega_2 +
r_+^2 d\chi^2$ which is precisely the flat metric on $R^{1,3} \times S^1$.

A scaling limit leading to Milne space will not be possible in Schwarzschild coordinates since these only cover the region outside the topological black hole horizon while, as we saw, the Milne spacetime arises from focusing near the singularity.

\subsection{Mass}

In asymptotically AdS spaces it is convenient to construct the charges of the spacetime (mass, angular momentum etc.) in terms of a boundary stress tensor \cite{vijayper,hensken}, since this formulation will agree with corresponding quantities in the dual field theory.   To compute masses, the time-time component of this stress tensor is integrated over the equal time boundary slice.   The computation  follows the methods in \cite{vijayper} and can be found in Appendix C.  (Also see the stress tensor computations in \cite{cai}.)  The masses we compute will be dimensionless because we are working with a dimensionless time.  The mass with respect to Schwarzschild time (\ref{schwarzschild}) is 
\begin{equation}
M_{{\rm topol}}^{{\rm Schw}} = - {\pi \over 8G} R_{{\rm AdS}}^2 r_+
  \cosh^2{r_+ t \over R_{{\rm AdS}}}
\Label{topschwmass}
\end{equation}
This is time dependent because $\partial_t$ in Schwarzschild coordinates is not a Killing direction.      Working in static coordinates (\ref{staticmet}) we can arrive at a notion of conserved mass, which is associated to a conserved energy measured by an inertial observer in the dual CFT.   This gives
\begin{equation}
M_{{\rm topol}}^{{\rm static}} = - { \pi \over 32 G}  R_{{\rm AdS}}^2 r_+
\Label{topstaticmass}
\end{equation}
Later we will compare these masses to those of the spacetime resulting from  the semiclassical decay.  Since the boundary of our spacetime is topologically different from that of global AdS space, a comparison between their masses is meaningless.

\section{Semiclassical instability}
 \label{sec:semiins}
 
 Following \cite{perry,wittenbubble,colemanbook}, a semiclassical decay in gravity is computed by the method of Euclidean bounces.  The idea  is to look for a saddlepoint to the Euclidean equations of motion that has the same boundary asymptotics as the Euclidean false vacuum.  If such a solution exists, and has a negative non-conformal mode (a fluctuation that decreases the Euclidean action), it provides an instanton for decay of the false vacuum.   The Lorentzian semiclassical description of the process is obtained by cutting open the Euclidean bounce on an equal time slice (say $t=0$) and then analytically continuing to Lorentzian signature.  The resulting expanding bubble simply replaces the $t>0$ part of the false vacuum solution.    This describes the sudden appearance of an expanding bubble in spacetime.  
 
 Tunneling in pure gravity has a somewhat different character from field theory since there is no potential in configuration space making it easy to identify the ``false" and ``true" vacua.    One important subtlety concerns the meaning of requiring that a solution in gravity satisfies a set of boundary conditions.   In gravity, because of reparametrization invariance it is not strictly meaningful to simply require that the boundary metric approaches a specified form.  The rate of approach is also crucial, to guarantee asymptotically flat (or AdS or dS) boundary conditions.   In addition, encoded in the rate of approach are the conserved charges of the spacetime.  Since we know that quantum tunneling cannot change the conserved charges, we should check that bounce solutions have the appropriate falloffs at infinity.   If such an analysis shows  that energy is not conserved, the tunneling process can only happen if it is accompanied by nucleation of other matter to make up the deficit.     Below we will use these methods to show that the topological black hole described in the previous section decays by tunneling to a bubble of nothing.

\subsection{Decay of the topological black hole}
 \label{eq:decay}

Following the discussion above, the bounce we seek is a Euclidean spacetime with the same asymptotic geometry as the Euclidean continuation of the topological black hole.   In Sec.~2.3 we showed that the latter is simply the thermal AdS geometry  the boundary of which is topologically $S^3 \times S^1$.    Another spacetime with the same boundary is the Euclidean AdS-Schwarzschild black hole.

The Lorentzian AdS-Schwarzschild black hole is
\begin{equation}
  ds^2 = -\left(1+\frac{r^2}{R_{\AdS}^2} -
  \frac{r_0^2}{r^2}\right)\,dt^2 +
  \left(1+\frac{r^2}{R_{\AdS}^2} -
  \frac{r_0^2}{r^2}\right)^{-1}\,dr^2 +
  r^2\,d\Omega_3^2~,
 \Label{eq:adsschwarzs}
\end{equation}
with a horizon at $r_h^2 = \frac{R_{\AdS}^2}{2}\left[-1 +
  \sqrt{1+\frac{4r_0^2}{R_{\AdS}^2}}\right]$.  
Its euclidean continuation $t \to -i\chi$ is
\begin{equation}
  ds^2 = \left(1+\frac{r^2}{R_{\AdS}^2} -
  \frac{r_0^2}{r^2}\right)\,d\chi^2 +
  \left(1+\frac{r^2}{R_{\AdS}^2} -
  \frac{r_0^2}{r^2}\right)^{-1}\,dr^2 +
  r^2\,d\Omega_3^2~.
 \Label{eq:eadsschwarzs}
\end{equation}
To avoid conical singularities we identify
\begin{equation}
  \chi \sim \chi + \frac{2\pi\,R_{\AdS}^2\,r_h}{2r_h^2 + R_{\AdS}^2}~.
 \Label{eq:period}
\end{equation}
Clearly, as $r \to \infty$ the boundary is topologically $S^3 \times S^1$.  Requiring \eqref{eq:eadsschwarzs} to also have the same asymptotic geometry as thermal AdS (or the euclidean topological black hole) relates the horizon radius $r_h$ to $r_+$, the scale of the circle in (\ref{scont}).
Either by matching periodicities, $\Delta \chi_{\textrm{topol.}} = 
\Delta \left( \frac{\chi_{\textrm{Eucl. Schw.}}}{R_{\AdS}}\right)$, or
by matching the ratio of the size of the $S^3$ and the $S^1$ at  infinity
\begin{equation*}
  \left. \frac{\textrm{Radius of } \Sph^1}{\textrm{Radius of }
  \Sph^3} \right|_{\textrm{topological}} = \left. 
  \frac{\textrm{Radius of} \Sph^1}{\textrm{Radius of } \Sph^3} 
  \right|_{\textrm{Schwarzschild}}, \quad \textrm{as } r\to \infty,
\end{equation*}
we learn that
\begin{equation}
 r_h = \frac{R_{AdS}^2}{4r_+} \left( 1 \pm \sqrt{1 -
 \frac{8r_+^2}{R_{AdS}^2}}\right).
\Label{eq:match}
\end{equation}
If the size of the circle at the horizon of the topological black hole
$(r_+)$ is larger than $(r_+)_{\text{max}}=\frac{R_{AdS}}{2\sqrt{2}}$,
there is no real solution to \eqref{eq:match}. This is  the well-known 
statement that AdS Schwarzschild black holes only exist for temperatures above
$T_0= \frac{R_{AdS}}{(r_+)_{\text{max}}}$.   Alternatively, the range of values for the periodicity of Euclidean time that are admitted by (\ref{eq:period}) has an upper bound.    

When $r_+ \leq (r_+)_{\text{max}}$ there
are two geometries (\ref{eq:eadsschwarzs}), the small and large AdS-Schwarzschild black holes, corresponding to the two solutions of (\ref{eq:match}).    These coincide when $r_+ = (r_+)_{\text{max}}$. The small AdS black holes have
negative specific heat, whereas large ones have positive specific
heat leading to very different thermodynamic properties.   For our purposes, the essential point is that the 
small\ euclidean AdS-Schwarzschild black hole has a non-conformal negative
mode \cite{perry,hawkingpage}, making it a candidate bounce.  The large black hole by constrast does not have a negative mode and cannot mediate a semiclassical instability despite having the same Euclidean asymptotics as our false vacuum.    Formally, in  a flat space limit analogous to the ones in Sec. 2.4 (i.e., $R_{{\rm AdS}} \to \infty$ with suitable rescalings of  coordinates) the small black hole goes over to the standard Schwarzschild spacetime while the large black hole grows to infinite size.  Thus it is natural to expect that in this limit, following \cite{wittenbubble}, the small AdS black hole will be one mediating a decay.

In summary, when $r_+ >(r_+)_{\text{max}}$, there is no bounce solution.  The topological black hole in this range of parameters is semiclassically stable.  In addition, the circle $\chi$ in the Euclidean Schwarzschild spacetime is contractible, and so fermions must be anti-periodic around this circle.   By contrast, fermions can be either periodic or anti-periodic around the $\chi$ direction of the topological black hole.  If they are periodic, this space is again stable against semiclassical decay for any value of $r_+$.  With anti-periodic boundary conditions, when $r_+ < (r_+)_{\text{max}}$ the small Euclidean AdS-Schwarzschild geometry gives rise to a bounce.

According to the general theory of semiclassical vacuum
decay, the false vacuum, i.e. the topological black hole in our
discussion, decays into a real on-shell lorentzian spacetime which
agrees with the euclidean bounce, i.e. the euclidean continuation
of the small AdS-Schwarzschild black hole, on a fixed hypersurface
that can be regarded as the ``origin'' of time $\tau=0$ after the
quantum tunneling process takes place.  Here, to use this prescription
we take the $\Sph^{3}$ sphere in
\eqref{eq:eadsschwarzs} and parameterize it as
\begin{equation}
  d\Omega_3^2 = d\theta^2 + \sin^2\theta\,d\Omega_2^2~.
\end{equation}
The hypersurface defined by $\theta=\frac{\pi}{2}$ will become
the aforementioned $\tau=0$ surface.  Consider the
analytic continuation $\theta \to \frac{\pi}{2} + i\tau$:
\begin{equation}
  ds^2 = \left(1+\frac{r^2}{R_{\AdS}^2}-\frac{r_0^2}{r^2}\right)\,d\chi^2 +
  \left(1+\frac{r^2}{R_{\AdS}^2}-\frac{r_0^2}{r^2}\right)^{-1}\,dr^2 
  -r^2\,d\tau^2 + r^2\,\cosh^2\tau\,d\Omega_2^2~.
 \Label{eq:topoadsbubble}
\end{equation}
This is simply the small bubble of nothing solution constructed in~\cite{vijaysimon}.  Because of the time dependence of the size of the boundary $S^2$, a coordinate independent notion of boundary time is given by the ratio of the size of the $S^2$ and the $S^1$ as $r \to \infty$.    This gives a way to unambiguously identify the instant when a transition occurs by matching this ratio computed in the topological black hole and the  bubble.  Although we have above chosen the $\tau = 0$ slice to perform such a matching, any other instant would in principle do.   Although the boundary times can be matched in this way, it is less clear how to match the slices in the bulk since a given boundary time can be extended in many ways into the bulk.

\paragraph{Localization: }
In IIB string theory $\ads{5}$ is always accompanied by some other five dimensional compactification manifold $\CM$.  The most symmetric situation occurs when $\CM$ is a 5-sphere.  In this case  it is expected that the small AdS-Schwarzschild black hole (which is smeared out over the $S^5$) is unstable to localization onto the $S^5$ in analogy with the Gregory-Laflamme instability for black strings \cite{greglaf}.\footnote{There is some evidence  that the Gregory-Laflamme instability produces inhomogeneous black strings rather than localized black holes (see~\cite{horowitzmaeda} and references thereto).   In analogy, this sort of instability of small AdS black holes might produce a horizon that is spread over the $S^5$ in an inhomogeneous way.}   The argument for the instability can be made on entropic grounds -- the  black hole localized in 10 dimensions will have a higher entropy than the 5d holes that we have considered \cite{simonamanda}.
In fact in the flat space limits described in Sec.~2.4 such a localization is necessary in order to reproduce the known instability of the flat Kaluza-Klein vacuum in $R^{8,1} \times S^1$ dimensions for which the bounce should be the 10 dimensional Schwarzschild spacetime~\cite{wittenbubble}.

\subsection{Energy conservation}

Since our false vacuum does not have any globally timelike Killing vectors, energy is not  conserved,  as we see from the time-dependent mass measured in global boundary coordinates (\ref{topschwmass}).   However, as we discussed, the boundary de Sitter geometry can also be written in the static coordinates appropriate to an inertial boundary observer, leading to the constant mass (\ref{topstaticmass}).  There is no contradiction between these two quantities, even though only one is time-dependent, because they are measuring the eigenvalues of different Hamiltonians.   We  require that during a quantum tunneling event energy must be conserved.  In the present context, this means that in static boundary time we ask whether the bubble of nothing and the topological black hole have equal masses.   In global boundary time, we should match the instant of tunneling as described above and then ask whether the masses coincide.

Again following the techniques in~\cite{vijayper,hensken} as reviewed in Appendix~C, one finds that using global boundary time (\ref{eq:topoadsbubble}) the small bubble of nothing has a mass\footnote{The masses we compute will be dimensionless because we are working with a dimensionless time.}
\begin{equation}
M_{bubble}  =  - \frac{\pi}{8G} r_h (2r_h^2 +R_{{\rm AdS}}^2) \cosh^2 \tau
\Label{smallbubblemass}
\end{equation}
This should be compared to global time topological black hole mass (\ref{topschwmass}).  The time  $r_+ t/R_{{\rm AdS}}$ in (\ref{topschwmass}) should be regarded as being the same as  $\tau$ in order to match the boundary geometries.     Then on any equal time slice $\tau = r_+ t/R_{{\rm AdS}}$ it is easy to show that $M_{{\rm  bubble}} < M_{{\rm topol}}^{{Schw}}$.   

For simplicity, consider the $t=\tau = 0$ surface.   Then,
using (\ref{eq:match}) for the small bubble of nothing we get
\begin{equation}
M_{{\rm bubble}}^{{\rm global}} =  -\frac{\pi}{8G}
 \frac{R_{{\rm AdS}}^6}{16r_+^3}
\left(1-\sqrt{1-\frac{8r_+^2}{R_{{\rm AdS}}^2}}\right)^2.
\end{equation}
It is easy to show that this is always less than the mass of the topological black hole in the range of $8r_+^2/R_{{\rm AdS}}^2 < 1$ where the bubble of nothing exists.  For simplicity, expanding the square root when this ratio is small we find that on the $t=\tau = 0$ surface
\begin{equation}
M_{{\rm bubble}}^{{\rm global}} = M_{{\rm topol}}^{{\rm Schw}} - \frac{\pi}{2G} r_+^3 - \frac{5\pi}{2G}
\frac{r_+^5}{R_{{\rm AdS}}^2} + ... < M_{{\rm topol}}^{{\rm Schw}} \, .
\end{equation}
Writing the de Sitter factor on the boundary in static coordinates as in (\ref{staticmet}) we find that the mass of the bubble measured by an inertial boundary observer is
\begin{equation}
M_{{\rm bubble}}^{{\rm static}} = -\frac{\pi}{32G} r_h \left(2r_h^2 + R_{{\rm AdS}}^2\right)  = -\frac{\pi}{32G}
\frac{R_{{\rm AdS}}^6}{16r_+^3}
\left(1-\sqrt{1-\frac{8r_+^2}{R_{{\rm AdS}}^2}}\right)^2.
\end{equation}
This is always less than the corresponding topological black hole mass (\ref{topstaticmass}) since both differ by a numerical factor from the measurement on the global $t=\tau=0$ slice.

Because our false vacuum has a higher mass than the bubble of nothing, the decay will only proceed if it is accompanied by the nucleation of sufficient energy to take up the mass differences computed above.   This excess energy can be in the form of a thermal gas of appropriate temperature.   
We can imagine dressing the bounce  solution with a spherically symmetric distribution of matter localized near the Euclidean origin and the backreaction of this matter will adjust the asymptotic falloff of the metric in the correct way to yield a space of the correct mass.    It is interesting to consider how the added matter would modify the time evolution of the Lorentzian bubble after it is nucleated.   Note that the large bubble of nothing has a mass that is less than that of the small bubble.  Therefore  there is no instanton for the spontaneous decay of this spacetime.  However one could imagine exciting matter in the large bubble background, providing the excess energy  that would then permit tunneling to the small bubble.

The decay of the flat space Kaluza-Klein vacuum did not have to contend with these energy conservation subtleties \cite{wittenbubble} even though the bubble of nothing there was constructed in a very similar way, namely by analytically continuing a Schwarzschild black hole.  This is because  in flat space the circle remained of constant size everywhere in the false vacuum.  Therefore,  the space only had four large dimensions at infinity and  the analytic continuation of a 5d Schwarzschild black hole had an asymptotic falloff in its metric ($\sim 1/r^2$)  that was too rapid to affect the mass of the 4d bubble.   Thus, both the false vacuum and the bubble of nothing had the same mass.   By contrast, in our case, the negative cosmological constant forces the circle in the false vacuum to grow in size towards infinity, leading to an asymptotically five dimensional geometry.  Hence, the analytic continuation of the 5d AdS-Schwarzschild black hole has a falloff in its metric that is sufficient to make a contribution to the mass.  This leads to a mass difference compared with the false vacuum with which we started.

\subsection{Euclidean actions of the relevant spaces}

\begin{figure}
\begin{center} 
\includegraphics{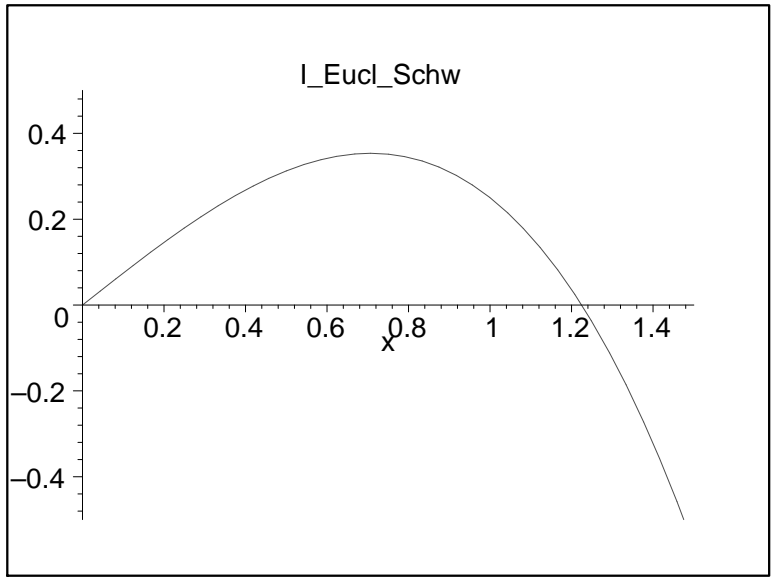}\label{f2}
\caption{Action of the Euclidean Schwarzschild black hole as a function of $x = r_h/R_{{\rm AdS}}$ where the periodicity of Euclidean time is set by in terms of $r_h$ by (\ref{eq:period}).    The small black holes have $x < 1/\sqrt{2}$  while the large black holes have $x>1/\sqrt{2}$.  We have set $3\pi^2 R_{{\rm AdS}}^3 / 16G =1$.}
\end{center}
\end{figure}

The decay rate of the false vacuum will depend on the difference in actions between the Euclidean false vacuum and the bounce solution \cite{colemanbook}.   To study this we compute the actions of these spaces using the counterterm method of \cite{vijayper}. The lorentzian actions are then 
given by
\begin{equation}
I = -\frac{1}{16\pi G} \int_M d^{5}x \sqrt{-g} \left( R - 
\frac{12}{R_{\AdS}^2} \right) - \frac{1}{8 \pi G}  \int_{\partial M} 
d^4x \sqrt{-\gamma} \Theta + \frac{1}{8\pi G} I_{ct}(\gamma),
\end{equation}
where $\gamma$ is the boundary metric and $\Theta$ is the trace of the 
extrinsic curvature on the boundary. $I_{ct}$ is the counterterm action 
added to obtain a finite stress tensor as in \cite{vijayper,hensken}.   To find the false vacuum action  we recall
that the euclidean continuation of the topological black hole is
thermal anti-de Sitter space (\ref{eq:etopbis}) with temperature $T = \frac{1}{\beta} =
\frac{R_{\AdS}}{r_+}$.   The Euclidean action is\footnote{We have chosen sign conventions so that action differences between spacetimes will have the same signs as the conventional background subtraction calculations in \cite{hawkingpage}.} 
\begin{equation}
I_{\textrm{Topol.}}^E = \frac{3\pi^2}{16 G} R_{\AdS}^2 r_+.
\end{equation}
The action of the bounce, namely the Euclidean Schwarzschild black hole  (\ref{eq:eadsschwarzs}), is
\begin{equation}
I_{\textrm{Eucl. Schw.}}^E = \frac{\pi^2 R_{\AdS}^3}{4 G} \frac{x}{1+2x^2} \left(
\frac{3}{4} + x^2 - x^4\right), 
\end{equation}
where $x = \frac{r_h}{R_{\AdS}}$.  The actions are displayed in Fig.~2.  It is easy to show that  for any given temperature (or Euclidean time periodicity) the small black hole has an action that is greater than that of the large black hole.   (At a given temperature there will be two solutions for $r_h$ corresponding to large and small black holes and the corresponding points in Fig.~\ref{f2} are at different heights.)   As discussed by Hawking and Page \cite{hawkingpage}, this implies that the large black holes dominate over the small ones in the contribution to the Euclidean path integral, or equivalently in the canonical ensemble for thermal AdS spacetimes.

Since we have a relation between the parameters $r_+$ and $r_h$ from the
matching of the boundaries, we can compare the actions of the false vacuum and the Euclidean Schwarzschild black holes with the same boundary (Fig.~3).  The action of the small Schwarzschild black hole ($x < 1/\sqrt{2}$) is always 
larger than the action of the false vacuum, but the relative sizes 
of the actions of the large black hole and our false vacuum depend  on the parameter $r_+$.   In the context of the canonical ensemble for global AdS space, this was interpreted as saying that at moderate temperatures the Euclidean path integral is dominated by thermal AdS, while at high temperatures the large black hole dominates.  (Recall that at low enough temperatures, the only saddlepoint is thermal AdS.)   In these discussions the small black hole never dominated.   However, in our context, where there is no thermal bath, the small black hole, which is known to have a negative mode, is the instanton mediating the decay of the false vacuum.  The action difference between the small black hole and the false vacuum then provides the decay rate of the latter, following \cite{perry,wittenbubble,colemanbook}.

\begin{figure}
\begin{center} 
\includegraphics{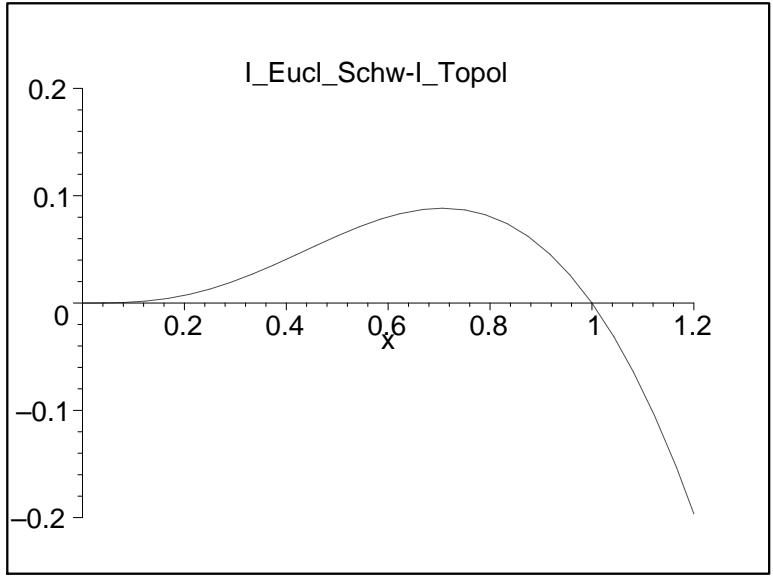}\label{f3}
\caption{Action difference between the Euclidean Schwarzschild black holes and the false vacuum as a function of $x = r_h/R_{{\rm AdS}}$.  The bounce solutions, corresponding to the small black holes, arise for $x < 1/\sqrt{2}$.  We have set $3\pi^2 R_{{\rm AdS}}^3 / 16G =1$.}
\end{center}
\end{figure}

%

\paragraph{Flat space limit: }

It is easy to show that the action difference between our false vacuum and our bounce diverges in the flat space limit after accounting for the fact that the 5d Newton constant and the 10d Newton constant are related as $G_5 \sim G_{10}/R_{{\rm AdS}}^5$.     This is consistent, because, as discussed in Sec.~3.2 in the presence of an additional $S^5$, we expect an instability for our bounce to localize onto the sphere.   In the flat-space limit we should expect this localization to reproduce the correct instanton.  Without it, our bounce becomes a five dimensional black sheet in the flat space limit.  
Another way of saying all this is that the expected existence of a localization instability tells us that our bounce, when embedded in $\ads{5} \times S^5$ must have additional negative modes that can reduce its action.   Condensing these additional negative modes will produce the true minimum action bounce solution which will be appropriately localized.

\section{Discussion}

\paragraph{Fluxbranes: } This paper described an instability of a particular boost orbifold of AdS space.   In fact, it is easy to see that our methods apply to a much wider class of spacetimes.   In flat space, fluxbranes are a famous class of unstable compactifications (see \cite{gibhor1,gibhor2} and references thereto).   These are quotients of Minkowski spacetime in which the identification of points is along a translation
plus a simultaneous rotation on a transverse space\footnote{There are
obvious generalisations with more than one rotation parameter.}
\begin{equation*}
  \xi = r_+\,\partial_x + \theta\,R_{ij}~,
\end{equation*}
where $R_{ij}$ stands for a rotation in the $ij$-plane.
Generically, the fluxbranes break supersymmetry and are well-known to decay semiclassically
\cite{gibhor1,gibhor2,brecher}.    But, using our previous discussions, it is possible to exhibit AdS orbifolds that reduce to the fluxbranes in scaling limits that generalize \eqref{eq:flatkk}.  Specifically, consider the quotient of AdS generated by a deformation of the Killing
vector \eqref{eq:boost} that was used to make the topological black hole:
\begin{equation*}
  \xi^\prime = \frac{r_+}{R_{{\rm AdS}}}(x_4 \partial_5 + x_5\partial_4)
  + \theta\,(x_i\partial_j - x_j\partial_i)~ .
  \Label{newquotient}
\end{equation*}
Here both $\{i,\,j\}$ stand for spacelike dimensions in $\RR^{2,4}$
transverse to $x^4$.  Taking the limit \eqref{eq:flatkk} while holding $\theta$ constant one recovers the fluxbranes.  Thus, one expects these AdS analogs of fluxbranes to decay semiclassically.   Moreover,
since the instanton in the fluxbrane decay involves an analytic
continuation of the Kerr black hole, it is natural to expect that 
rotating black holes in AdS will play a similar role for quotients by (\ref{newquotient}).    In other
words, the bubbles of nothing constructed in \cite{vijaysimon} out of
the Kerr AdS black holes should describe the semiclassical decay of AdS-fluxbranes.

\paragraph{Field theory dual: }    In fact many instabilities of flat space, such as the closed string tachyon condensation of (e.g., \cite{adams}), can be embedded into AdS space using the kind of reasoning described above.   The reason to do this is the hope that a dual field theory description of catastrophic closed string instabilities can be found using the AdS/CFT correspondence.    In the example of the present paper, the dual field theory should be $\CN=4$ SU(N) super-Yang-Mills at large $N$ on three dimensional de Sitter space times a circle.  We know that this is  the dual because, as we showed, our space can be recovered as the near horizon limit of D3-branes whose worldvolume fills the Milne orbifold.\footnote{Of course this near horizon limit only gives a patch of the full spacetime, just as the usual scaling limit of flat D3-branes gives the Poincar\'e patch of global AdS.}   In string theory, as we have discussed, there are two instabilities of interest.  First, the topological black hole can tunnel into a bubble of nothing and second, the bubble that we displayed should have a perturbative instability to localization on the $S^5$.  We  expect that Yang-Mills theory has a suitable effective potential with a false vacuum describing the topological black hole.  Tunneling out of this false vacuum should describe the appearance of the bubble of nothing.  Since the bubble we describe is delocalized on $S^5$, the tunneling process should involve singlets of the $SO(6)$ R-symmetry of the Yang-Mills theory.  By contrast, the perturbative localization instability breaks $SO(6)$ and should involve rolling in a direction of field space that breaks R-symmetry.   In all of this it is crucial that the Yang-Mills theory breaks supersymmetry, and that in addition the fermions are anti-periodic around the circle in the background geometry.      What is more, since the flat space limit leads to the classic decay of the Kaluza-Klein vacuum in 10 dimensions, we  expect that the analogous scaling in the field theory leads to a new description of the decay of flat space.  Since the limit involves focusing on the deep interior of the spacetime, by the UV/IR correspondence in AdS/CFT, we should expect only the deep infrared physics of the Yang-Mills theory to contribute.  Since the theory is defined on a compact space, it is possible that the relevant infrared limit results in a matrix model reduction.      The first step to achieving these results is to identify the correct degrees of freedom for which an effective potential should be computed.     To this end, it is intriguing that the Euclidean continuation of our CFT setup is precisely the thermal Yang-Mills theory (on $S^3 \times S^1$) that is relevant for the Hawking-Page transition; we are in effect considering a novel analytic continuation and dynamical interpretation of the free energy that appears in the thermal setting.  Recent attempts to compute this free energy have focused on the Polyakov loop as an order parameter for the Hawking-Page transition \cite{sundborg,ofer,hong}.   This cannot be the complete story for us, since the Polyakov loop is a singlet under the R-symmetry and cannot account for the localization effect that is important here.  In any case, these free energies are computed at weak coupling and, in the absence of supersymmetry, it is not obvious that the results continue to strong coupling in a simple way.  Regardless, this is an intriguing possibility for future work.\footnote{The  model for the Hawking-Page transition that is developed in~\cite{hongnew} might well be useful.}

\section*{Acknowledgments}
We thank Ofer Aharony, Micha Berkooz, Jan de Boer, Gary Horowitz, Hong Liu, Rob Myers, Asad Naqvi, Eliezer Rabinovici, Simon Ross and Kostas Skenderis for useful conversations and email exchanges. The work of JS, VB and KL is supported by the DOE under grant DE-FG02-95ER40893, 
 by the NSF under grant PHY-0331728 and OISE-0443607.

\appendix

\section{Flat limits of AdS quotients}
\label{sec:forb}

It is well-known that there exist two inequivalent Inon\"u-Wigner
contractions of the conformal algebra $\fso(2,p)$. One 
contraction gives rise to the Poincar\'e algebra $\fso(1,p)\ltimes
\RR^{1,p}$; the second to the symmetry algebra of some suitable
pp-wave. At a purely algebraic level,
these contractions correspond to non-trivial scaling limits of the
generators of a given algebra, keeping its structure constants
finite. Since these algebras can be realised as isometry algebras
of spacetime metrics, the corresponding contractions can be realised
geometrically. For the pp-wave geometry this corresponds to the so
called Penrose limit of AdS, whereas for Minkowski spacetime, it corresponds
to the flat limit of AdS.

Let us review the latter.   Given global $\AdS_{p+1}$
\begin{equation*}
  g_{\AdS_{p+1}} = R^2_{\AdS}\left[\cosh^2\rho\,d\tau^2 + d\rho^2 + 
  \sinh^2\rho\,g_{\Sph^{p-1}}\right]~,
\end{equation*}
where $g_{\Sph^{p-1}}$ stands for the metric of a unit radius
$(p-1)$-sphere, the flat limit corresponds to sending
$R_{\AdS}\to\infty$ while keeping the metric finite. This is achieved by
\begin{equation*}
  R_{\AdS}\to \infty ~~,~~
  \tau \to \frac{t}{R_{\AdS}}~~,~~
    \rho\to\frac{r}{R_{\AdS}}~~~ t,r\,\,\text{fixed}
\end{equation*}  

Since we are interested in studying the behaviour of the symmetry
algebra under such limits, it is convenient to describe 
$\AdS_{p+1}$ in terms of an embedded hyperboloid in $\RR^{2,p}$
\begin{equation*}
  -(x^1)^2 - (x^2)^2 + \sum_{i=3}^{p+2} (x^i)^2 = R_{\AdS}^2~,
\end{equation*}
where the isometry group $\SO(2,p)$ acts linearly. The connection
among these cartesian coordinates and global coordinates is
\begin{eqnarray*}
  x^1 &=& R_{\AdS}\,\cosh\rho\,\cos\tau ~,\\
  x^2 &=& R_{\AdS}\,\cosh\rho\,\sin\tau ~,\\
  x^i &=& R_{\AdS}\,\sinh\rho\,\hat{x}^i \quad i = 3,\dots ,p+2
\end{eqnarray*}
where $\{\hat{x}^i\}$ parameterise a unit $(p-1)$-sphere. Given the
one--to--one correspondence among elements of the algebra $\fso(2,p)$
and the two forms $e_{mn}=x_m\partial_n-x_n\partial_m$, it is evident
that given our choice of $\{x^1,\,x^2\}$ plane in the above
parameterisation,\footnote{There exist others related by an $\fso(2)$
transformation rotating both timelike axes.} the set of generators
$\{e_{mn}\}$ behaves as follows in the flat limit defined above:
\begin{equation}
  e_{12}\to \partial_t~~,~~  e_{1i}\to \partial_i ~~,~~
  e_{2i}\to t\partial_i + x^i\partial_t ~~,~~
  e_{ij}\to e_{ij}~,
 \label{eq:dictio}
\end{equation}
after taking into account appropiate rescalings. The above
transformation is a geometric realisation of the abstract algebraic
contraction.

Given a quotient of AdS generated by the action of the Killing vector
$\xi$ associated with the form $e_{ab}$, the isometries $\{\xi_a\}$ 
preserved by the quotient space are those that commute with the Killing
vector $([\xi,\,\xi_a]=0)$. Clearly, if $\xi$ commutes with the
$\fso(2)$ transformation rotating both timelike axis, there will exist
a single flat limit. On the contrary, if the generator of the quotient
does not commute with such ``timelike'' rotations, there will be two
inequivalent flat limits, as seen in the present work.  To identify the quotient
of Minkowski spacetime that one would be left with after the flat
limit, one just has to identify the behaviour of the generator $\xi$
under the corresponding flat limit. Thus, for example, consider the
generator of the topological black hole in five dimensions
\begin{equation*}
  \xi = x^1\partial_6 + x^6\partial_1~.
\end{equation*}
The first flat limit corresponds to sending $x^2\to t$, and $x^1\to
R_{\AdS}$, as we did above. Thus, $\partial_1$ becomes a trivial
operator, and the finite limit of $\xi$ becomes a spacelike
translation. The corresponding quotient is the one giving rise to the
Kaluza--Klein vacuum. The second flat limit is inequivalent to
the first one since the topological black hole breaks the symmetry of
rotation among the two time axis in $\RR^{2,4}$.  It is achieved by $x^1\to
t$ and $v\to R_{\AdS}$. In this case, $\xi$ becomes a boost generator
of the Lorentz group $\SO(1,4)$, giving rise to the boost orbifold of
Minkowski spacetime in five dimensions.

The same discussion would apply for any generator in $\SO(2,p)$ in any
dimension. Given the classification of all inequivalent abelian
quotients of $\AdS_{p+1}$ \cite{paper5.1}, it is fairly simple to
establish a connection between their flat limits and the
classification of abelian quotients of Minkowski spacetime in
arbitrary dimension \cite{paper1} by using the dictionary
\eqref{eq:dictio} and identifying the conjugacy class to which they
belong. We summarise the relation among the building blocks of the
different quotients one could consider before and after the flat limit in
table \ref{tab:flimitblocks}.

\begin{table}[h!]
  \centering
  \setlength{\extrarowheight}{3pt}
  \renewcommand{\arraystretch}{1.3}
    \begin{tabular}{|>{$}l<{$}|>{$}l<{$}|}\hline
      \multicolumn{1}{|c|}{Minkowski} & \multicolumn{ 1}{c|}{AdS} \\
      \hline\hline
      \CC/\ZZ_n & \be_{ij} \\
      \RR/\ZZ = S^1 & \be_{1i} \\
      \RR^{1,1}/\ZZ & \be_{1i} \\
      \RR_t/\ZZ & \be_{12} \\
      \RR^{1,2}/\ZZ & \be_{13} - \be_{34} \\
      \hline
    \end{tabular}
  \vspace{8pt}
  \caption{Dictionary between generators of abelian quotients in AdS and
their corresponding flat limits, giving rise to conical singularities
$(\CC/\ZZ_n)$, spacelike and timelike circles $(\RR/\ZZ,\,\RR_t/\ZZ)$,
the boost orbifold $(\RR^{1,1}/\ZZ)$ and the null-rotation orbifold 
$(\RR^{1,2}/\ZZ)$. Linear combinations of the latter can give rise to
fluxbranes and nullbranes, with their corresponding images in AdS.} 
  \label{tab:flimitblocks}
\end{table}

\section{Brane probes}
\label{sec:probe}

In this appendix, we study the force felt by a brane in the presence of the topological black hole.
We use a coordinate system for the topological black hole that only covers the region outside the horizon:
\begin{equation*}
  y^\mu = \tanh^{1/2}\rho\,x^\mu\,, \quad x^\mu
  x^\nu\,\eta_{\mu\nu}=1~.
\end{equation*}
Thus $\{x^\mu\}$ parameterises de Sitter space in three dimensions
whereas $\rho\in (0,\,\infty)$. The metric
is
\begin{multline}
  g_{\text{top}} = \frac{4\,R_{\AdS}^2}{(1-\tanh\rho)^2}\,\left\{
  \tanh\rho\,g_{dS_3} + 
  \frac{1}{4\,\sinh\rho\,\cosh^3\rho}\,d\rho^2\right\} \\
  + r_+^2\,\left(\frac{1+\tanh\rho}{1-\tanh\rho}\right)^2\,d\chi^2~.
\end{multline}
Whereas the dilaton is constant, the five-form field strength
inn the AdS directions is
\begin{equation}
  F_{(5)} =
  -4R_{\AdS}^3\,r_+\,\frac{(1+\tanh\rho)^2}{(1-\tanh\rho)^4}\,\tanh\rho\,
  d\rho\,\wedge \dvol\,dS_3\wedge d\chi~,
\end{equation}
from which we extract the four form potential 
\begin{equation}
  C_{(4)} = -2 R_{\AdS}^3\,r_+\,\frac{\tanh^2\rho}{(1-\tanh\rho)^4}\,\dvol\,dS_3\wedge
  d\chi~,
\end{equation}
In these coordinates, it is natural to consider D3-branes
whose worldvolumes are $dS_3\times S^1$, located at a fixed radial
coordinate $\rho$. 
One can then compute the force felt by these
D-branes as a function of the location $\rho$. 
A straightforward calculation shows that the associated potential is
\begin{multline}
  V_\pm(\rho) =
  2T_{D3}\,R_{\AdS}^3\,r_+\,\sqrt{-\text{det}\,g_{dS_3}}\,
  \frac{\tanh\rho}{(1-\tanh\rho)^4}\\
  \left\{4\tanh^{1/2}\rho\,(1+\tanh\rho) \pm \tanh\rho\right\}~,
\end{multline}
Here $\pm$ refers to the potential felt by branes and anti-branes respectively.
In both cases, the probe will fall into
the horizon if released, indicating that these branes are not stable.

Note that the probe branes described above are not the same as the branes filling the Milne orbifold whose near-horizon limit gives (a patch of) the topological black hole.   In the coordinates used in this section,  those latter branes would have radial velocity, i.e.
$d\rho/dt\neq 0$.

\subsection{Bubble probe}
 \label{sec:bprobe}

Branes  can also probe the bubble of nothing  spacetime
\eqref{eq:topoadsbubble}. The relevant four-form potential
components are 
\begin{equation}
  C_{(4)} = -\frac{1}{8} R_{\AdS}^{-1}\,r^4\, \dvol dS_3\times d\chi~.
\end{equation}
As before, we can compute the potential that a D3-brane feels when
located at a fixed $r$, the answer being
\begin{equation}
  V_\pm (r) = T_{D3}\,r^3\,\sqrt{-\text{det}\,g_{dS_3}}\,
  \left\{\sqrt{1+\frac{r^2}{R_{\AdS}^2}- \frac{r_0^2}{r^2}} \pm
  \frac{1}{8}\,\frac{r}{R_{AdS}}\right\}~,
\end{equation}
where the minus would stand for an antiD3-brane.  In the physical region $r > r_h$,
 the potential felt by a D3-brane grows as a function of
$r$, which means that the probe will fall towards the bubble wall if released.
The same conclusion is reached for anti-D3-branes.

\section{Mass computations}
We will follow the techniques in \cite{vijayper} to compute the mass
of the topological black hole.  First  identify $N^2(r) = R_{{\rm
AdS}}^2/(r^2 - r_+^2)$ as the radial lapse function in the
Schwarzschild metric (\ref{schwarzschild}).  Defining $\tau = r_+ t /
R_{{\rm AdS}}$, this metric is
\begin{equation}
ds^2 = N(r)^2 dr^2 + (r^2-r_+^2)\frac{R_{{\rm AdS}}^2}{r_+^2} \left( -d\tau^2
+ \cosh^2 \tau (d\theta^2 + \sin^2 \theta d\varphi^2)\right) + r^2
d\chi^2.
\end{equation}
The middle terms (multiplied by $(r^2 - r_+^2)$) describe the 3d de Sitter spacetimes on each radial slice.   These geometries can be written in coordinates appropriate to inertial de Sitter observers (see, e.g., the review \cite{accreview}) giving\begin{equation}
ds^2 = N^2 dr^2 + \underbrace{\frac{R_{{\rm AdS}}^4}{r_+^2N^2} \left(
-(1-\rho^2) dt^2 + \frac{1}{1-\rho^2} d\rho^2 + \rho^2
d\theta^2\right) + r^2 d\chi^2}_{\gamma_{ij} = \textrm{boundary
metric}},
\end{equation}
where $0\le\rho<1$ and $0\le\theta\le 2\pi$.   This metric does not cover the entire boundary geometry, but is manifestly static so that the eigenvalue of $\partial_t$ will be conserved.

According to \cite{vijayper} the boundary stress tensor at a position $r$ is given terms of the extrinsic curvature of the boundary ($\Theta_{ij}$) and its intrinsic Einstein tensor ($G_{ij}$) as:
\begin{equation}
T_{ij} = {1 \over 8\pi G} \left(\Theta_{ij} - \Theta \gamma_{ij} -
\frac{3}{R_{{\rm AdS}}} \gamma_{ij} - \frac{R_{{\rm AdS}}}{2} G_{ij} \right)
\end{equation}
In our case the extrinsic curvature is
\begin{equation}
\Theta_{ij} = -\frac{1}{2N} \gamma_{ij,r}, \quad (i,j = t, \rho,
\theta, \chi) \, .
\end{equation}
We then find the boundary stress tensor in static coordinates on a fixed $r$ surface:
\begin{eqnarray}
T_{tt} & = & -\frac{R_{{\rm AdS}}\,r_+^2}{64\pi G} \frac{(1-\rho^2)}{r^2} + \mathcal{O}(r^{-4}), \\
T_{\rho\rho} & = & \frac{R_{{\rm AdS}}\,r_+^2}{64\pi G} \frac{1}{r^2(1-\rho^2)} + \mathcal{O}(r^{-4}), \\
T_{\theta\theta} & = & \frac{R_{{\rm AdS}}\,r_+^2}{64\pi G} \frac{\rho^2}{r^2} + \mathcal{O}(r^{-4}), \\
T_{\chi\chi} & = & - \frac{3}{64\pi G} \frac{r_+^4}{R_{{\rm AdS}}\,r^2} +
\mathcal{O}(r^{-4}).
\end{eqnarray}
We choose a spacelike surface $\Sigma$ on the boundary
$\partial M$: $t = const$. The metric on $\Sigma$ is 
\begin{equation}
ds^2_{\Sigma} = \frac{R_{{\rm AdS}}^4}{r_+^2N^2} \frac{1}{1-\rho^2} d\rho^2 +
\frac{R_{{\rm AdS}}^4}{r_+^2N^2} \rho^2 d\Theta^2 + r^2 d\chi^2.
\end{equation}
Let $u_{\mu} = \tilde{N} \delta_{\mu t} \equiv \frac{R_{{\rm AdS}}^2}{r_+N}
\sqrt{1-\rho^2} \delta_{\mu t} $ be a unit vector in $\partial M$
normal to $\Sigma$. Then the mass of the spacetime is
\begin{equation}
M = \int_{\Sigma} d^3 x \sqrt{\sigma} \tilde{N} u^{\mu} u^{\nu}
T_{\mu\nu},
\end{equation}
where $\sigma = \det(g_{\Sigma})$. This gives
\begin{equation}
M_{{\rm topol}}^{{\rm static}} = -\frac{\pi}{32G} r_+\,R_{{\rm AdS}}^2.
\end{equation}
A similar computation in Schwarzschild coordinates (\ref{schwarzschild}) give:
\begin{equation}
M_{{\rm topol}}^{{\rm schw}} =
 -\frac{\pi}{8G} R_{{\rm AdS}}^2 r_+ \cosh^2 \frac{r_+t}{R_{{\rm AdS}}}
\end{equation}

\providecommand{\href}[2]{#2}\begingroup\raggedright


\begin{thebibliography}{10}

\bbibitem{perry} 
M. ~Perry, ``Instabilities in gravity and supergravity'', In:
{\em Superspace and supergravity: Proceedings of the Nuffield Workshop}, 
Cambridge University Press, 1981.

\bbibitem{wittenbubble}
E.~Witten, ``Instability of the kaluza-klein vacuum,'' {\em Nucl.
Phys.} {\bf
  B195} (1982)
481.

\bbibitem{birm}
D.~Birmingham and M.~Rinaldi,
``Bubbles in anti-de Sitter space,''
Phys.\ Lett.\ B {\bf 544}, 316 (2002)
[arXiv:hep-th/0205246].


\bbibitem{vijaysimon}
V.~Balasubramanian and S.~F. Ross, ``The dual of nothing,'' {\em
Phys. Rev.}
  {\bf D66} (2002) 086002,
\href{http://www.arXiv.org/abs/hep-th/0205290}{{\tt
hep-th/0205290}}.


\bbibitem{biswas}
A.~Biswas, T.~K.~Dey and S.~Mukherji,
``R-charged AdS bubble,''
arXiv:hep-th/0412124.



\bbibitem{dine}
M.~Dine, P.~J.~Fox and E.~Gorbatov,
``Catastrophic decays of compactified space-times,''
JHEP {\bf 0409}, 037 (2004)
[arXiv:hep-th/0405190].



\bibitem{clean}
O.~Aharony, M.~Fabinger, G.~T.~Horowitz and E.~Silverstein,
``Clean time-dependent string backgrounds from bubble baths,''
JHEP {\bf 0207}, 007 (2002)
[arXiv:hep-th/0204158].



\bbibitem{simongeorgina}
S.~F.~Ross and G.~Titchener,
``Time-dependent spacetimes in AdS/CFT: Bubble and black hole,''
arXiv:hep-th/0411128.




\bbibitem{banados}
M.~Banados, A.~Gomberoff, and C.~Martinez, ``Anti-de sitter space
and black
  holes,'' {\em Class. Quant. Grav.} {\bf 15} (1998) 3575--3598,
\href{http://www.arXiv.org/abs/hep-th/9805087}{{\tt
hep-th/9805087}}.



\bbibitem{BTZ}
M.~Banados, M.~Henneaux, C.~Teitelboim and J.~Zanelli,
``Geometry of the (2+1) black hole,''
Phys.\ Rev.\ D {\bf 48}, 1506 (1993)
[arXiv:gr-qc/9302012];~~~
M.~Banados, C.~Teitelboim and J.~Zanelli,
``The Black hole in three-dimensional space-time,''
Phys.\ Rev.\ Lett.\  {\bf 69}, 1849 (1992)
[arXiv:hep-th/9204099].



\bbibitem{paper5.2} J. ~Figueroa-O'Farrill, O. ~Madden, S. ~Ross and J. ~Simon,
``Quotients of $AdS_{p+1} \times S^q$: causally well-behaved spaces
and black holes'', \href{http://www.arxiv.org/abs/hep-th/0402094}{{\tt
hep-th/0402094}}.

\bbibitem{accreview}
V.~Balasubramanian,
``Accelerating universes and string theory,''
Class.\ Quant.\ Grav.\  {\bf 21}, S1337 (2004)
[arXiv:hep-th/0404075].

\bbibitem{paper1}
J.~Figueroa-O'Farrill and J.~Simon, ``Generalized supersymmetric
fluxbranes,''
  {\em JHEP} {\bf 12} (2001) 011,
\href{http://www.arXiv.org/abs/hep-th/0110170}{{\tt
hep-th/0110170}}.

\bbibitem{paper5.1}
J.~Figueroa-O'Farrill and J.~Simon, ``Supersymmetric kaluza-klein
reductions of
  ads backgrounds,''
\href{http://www.arXiv.org/abs/hep-th/0401206}{{\tt
hep-th/0401206}}.

\bbibitem{joanAdS}
J.~Simon,
``Null orbifolds in AdS, time dependence and holography,''
JHEP {\bf 0210}, 036 (2002)
[arXiv:hep-th/0208165].

\bbibitem{simonowen}
O.~Madden and S.~F.~Ross,
Phys.\ Rev.\ D {\bf 70}, 026002 (2004)
[arXiv:hep-th/0401205].

\bbibitem{cai}
R.~G.~Cai,
``Constant curvature black hole and dual field theory,''
Phys.\ Lett.\ B {\bf 544}, 176 (2002)
[arXiv:hep-th/0206223].

\bbibitem{vijayper} 
V. ~Balasubramanian and P. ~Kraus, ``A Stress Tensor For 
Anti-de Sitter Gravity,'' {\em Comm. Math. Phys.} {\bf 208} (1999) 413--428,
\href{http://www.arxiv.org/abs/hep-th/9902121}{{\tt hep-th/9902121}}.

\bbibitem{hensken}
M.~Henningson and K.~Skenderis,
``The holographic Weyl anomaly,''
JHEP {\bf 9807}, 023 (1998)
[arXiv:hep-th/9806087];~~~
M.~Henningson and K.~Skenderis,
``Holography and the Weyl anomaly,''
Fortsch.\ Phys.\  {\bf 48}, 125 (2000)
[arXiv:hep-th/9812032];~~~
K.~Skenderis,
``Asymptotically anti-de Sitter spacetimes and their stress energy  tensor,''
Int.\ J.\ Mod.\ Phys.\ A {\bf 16}, 740 (2001)
[arXiv:hep-th/0010138].
m




\bbibitem{colemanbook}
S.~R.~Coleman,
``Aspects of Symmetry,'' Selected Erice Lectures, chapter 7, section
6, Cambridge University Press, 1985.

\bbibitem{hawkingpage} 
S. ~Hawking and D. ~Page, ``Thermodynamics of Black Holes
in Anti-De Sitter Space'', {\em Comm. Math. Phys.} {\bf 87} (1983) 577.

\bbibitem{greglaf}
R.~Gregory and R.~Laflamme,
``Black strings and p-branes are unstable,''
Phys.\ Rev.\ Lett.\  {\bf 70}, 2837 (1993)
[arXiv:hep-th/9301052];~~
R.~Gregory and R.~Laflamme,
``The Instability of charged black strings and p-branes,''
Nucl.\ Phys.\ B {\bf 428}, 399 (1994)
[arXiv:hep-th/9404071].

\bbibitem{horowitzmaeda}
G.~T.~Horowitz and K.~Maeda,
``Fate of the black string instability,''
Phys.\ Rev.\ Lett.\  {\bf 87}, 131301 (2001)
[arXiv:hep-th/0105111];~~~
G.~T.~Horowitz and K.~Maeda,
``Inhomogeneous near-extremal black branes,''
Phys.\ Rev.\ D {\bf 65}, 104028 (2002)
[arXiv:hep-th/0201241].

\bbibitem{simonamanda}
A.~W.~Peet and S.~F.~Ross,
``Microcanonical phases of string theory on AdS(m) x S(n),''
JHEP {\bf 9812}, 020 (1998)
[arXiv:hep-th/9810200].

\bbibitem{gibhor1}
F.~Dowker, J.~P. Gauntlett, G.~W. Gibbons, and G.~T. Horowitz,
``The decay of
  magnetic fields in kaluza-klein theory,'' {\em Phys. Rev.} {\bf D52} (1995)
  6929--6940,
\href{http://www.arXiv.org/abs/hep-th/9507143}{{\tt
hep-th/9507143}}.

\bbibitem{gibhor2}
F.~Dowker, J.~P. Gauntlett, G.~W. Gibbons, and G.~T. Horowitz,
``Nucleation of
  $p$-branes and fundamental strings,'' {\em Phys. Rev.} {\bf D53} (1996)
  7115--7128,
\href{http://www.arXiv.org/abs/hep-th/9512154}{{\tt
hep-th/9512154}}.

\bbibitem{brecher}
D.~Brecher and P.~M. Saffin, ``Decay modes of intersecting
fluxbranes,'' {\em
  Phys. Rev.} {\bf D67} (2003) 125013,
\href{http://www.arXiv.org/abs/hep-th/0302206}{{\tt
hep-th/0302206}}.

\bbibitem{adams}
A.~Adams, J.~Polchinski and E.~Silverstein, ``Don't panic! closed
string  tachyons in ale space-times,'' {\em JHEP} {\bf 10} (2001) 029,
\href{http://www.arXiv.org/abs/hep-th/0108075}{{\tt
hep-th/0108075}}



\bbibitem{sundborg}
B.~Sundborg,
``The Hagedorn transition, deconfinement and N = 4 SYM theory,''
Nucl.\ Phys.\ B {\bf 573}, 349 (2000)
[arXiv:hep-th/9908001];~~~
P.~Haggi-Mani and B.~Sundborg,
``Free large N supersymmetric Yang-Mills theory as a string theory,''
JHEP {\bf 0004}, 031 (2000)
[arXiv:hep-th/0002189].



\bbibitem{ofer}
O.~Aharony, J.~Marsano, S.~Minwalla, K.~Papadodimas and M.~Van Raamsdonk,
``The Hagedorn / deconfinement phase transition in weakly coupled large N
gauge theories,''
arXiv:hep-th/0310285.

\bbibitem{hong}
H.~Liu,
``Fine structure of Hagedorn transitions,''
arXiv:hep-th/0408001.

\bbibitem{hongnew}
Luis Alvarez-Gaum\'e, C\'esar G\'omez,  Hong Liu, and Spenta Wadia, 
``Finite temperature effective action, AdS$_5$ black holes, and $1/N$ 
expansion,'' to appear.




\end{thebibliography}
\end{document}